\documentclass[prl,twocolumn,preprintnumbers,amsmath,amssymb]{revtex4-1}
\usepackage{graphicx}
\usepackage{dcolumn}
\usepackage{amsmath,amsbsy,amssymb,scalefnt}
\usepackage{bm}
\usepackage{epstopdf}
\usepackage[T1]{fontenc}
\usepackage{amsfonts}
\usepackage{textcomp}
\usepackage{mathtools}
\usepackage{hyperref}
\usepackage{physics}
\hypersetup{ pdfnewwindow=true, colorlinks=true, linkcolor=blue, anchorcolor=blue, citecolor=blue, filecolor=blue, menucolor=blue, urlcolor=blue}

\begin{document}

\title{Light-induced dissipationless states in magnetic topological insulators\\with hexagonal warping}

\date{\today}
\author{Mohammad Shafiei}
\affiliation{Department of Physics \& NANOlight Center of Excellence, University of Antwerp, Groenenborgerlaan 171, B-2020 Antwerp, Belgium}
 
\author{Milorad V. Milo\v{s}evi\'c}
\email{milorad.milosevic@uantwerpen.be}
\affiliation{Department of Physics \& NANOlight Center of Excellence, University of Antwerp, Groenenborgerlaan 171, B-2020 Antwerp, Belgium}
\date{\today}

\begin{abstract}

Magnetic impurities in topological insulators (TIs) induce backscattering via magnetic torque, unlike pristine TIs where spin-orbit locking promotes dissipationless surface states. Here we reveal that one can suppress that unwanted backscattering and dissipation in magnetic TIs using high-frequency linearly polarized light (LPL). By carefully considering the hexagonal warping of the Fermi surface of the TI, we demonstrate how the coupling between Dirac surface states and LPL can effectively reduce backscattering on magnetic dopants, enhance carrier mobility and suppress resistance, even entirely. These findings open up avenues for designing ultra low-power sensing and spintronic technology.
\end{abstract}
\maketitle

\paragraph{Introduction} Topological insulators (TIs) are a class of materials with unique electronic properties, characterized by bulk insulating behavior and topologically protected metallic surface states~\cite{fu2007topological,hasan2010colloquium}. These surface states exhibit Dirac-cone-like energy dispersion and spin-momentum locking~\cite{qi2011topological,hasan2011three}, rendering them robust against backscattering in the absence of spin-flip processes~\cite{tian2017property}. This intriguing characteristic makes TIs highly promising for applications in spintronics and dissipationless electronics~\cite{shun2018topological}.

\begin{figure}[b]
    \centering
     \includegraphics[width=0.95\linewidth]{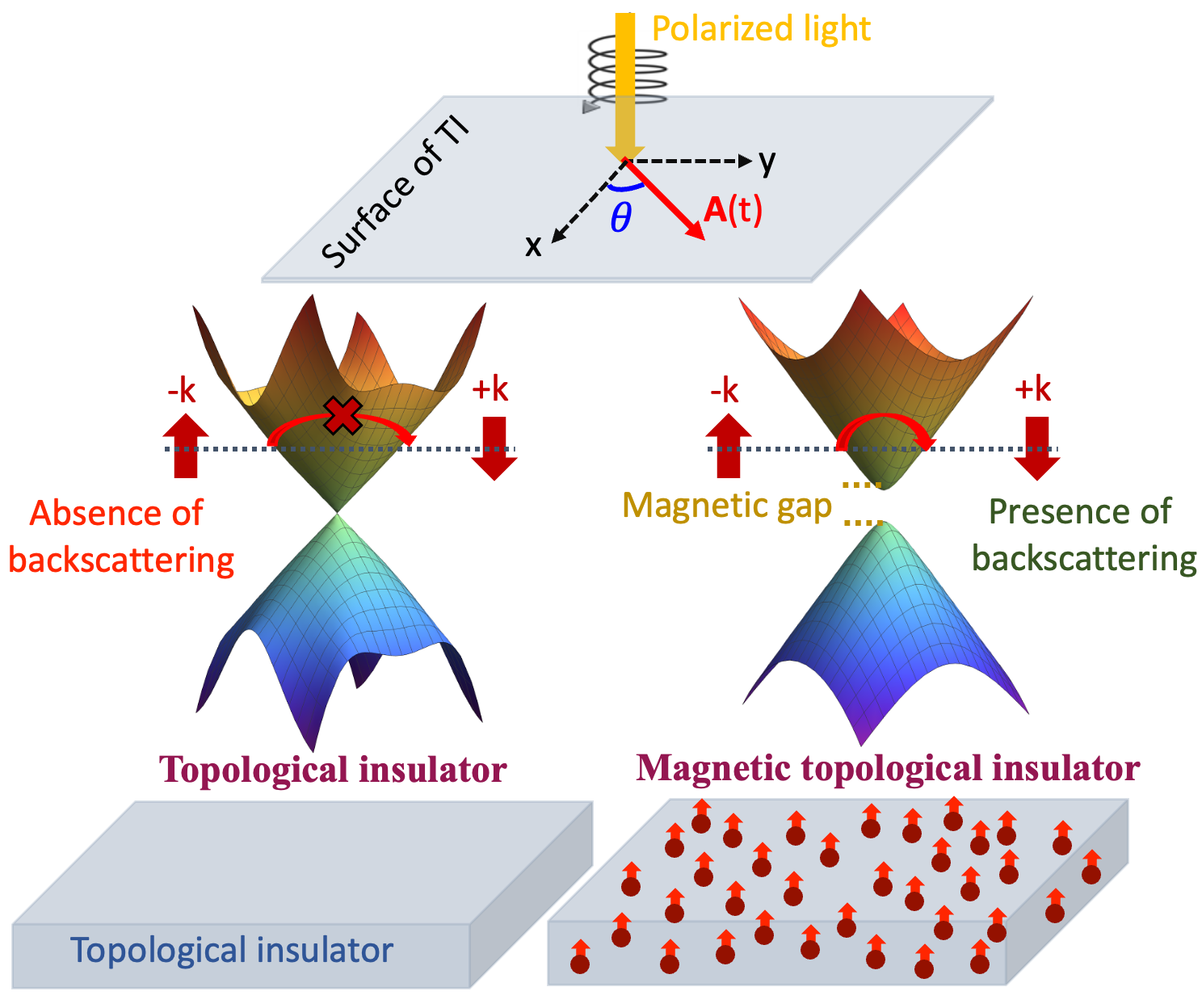}
    \caption{Schematic illustration of backscattering in pristine and magnetic TIs. In pristine TIs, helical surface states are protected from backscattering due to TRS. In magnetic TIs, TRS is broken, leading to backscattering and destruction of helical surface states. The top panel illustrates the illumination with high-frequency LPL, as used in this work, incident perpendicular to the TI surface, with a polarization angle $\theta$ relative to $x$-axis.}
    \label{fig:schematic_LPL}
\end{figure}

While three-dimensional TIs were initially proposed in time-reversal invariant systems, in magnetic TIs an exchange gap arises in the Dirac band dispersion (as shown schematically in Fig.~\ref{fig:schematic_LPL}) due to the presence of magnetization or, equivalently, a broken time-reversal symmetry (TRS)~\cite{tokura2019magnetic}.
In recent years, magnetic TIs and unique quantum states within these materials, such as quantum anomalous Hall states~\cite{chang2013experimental}, axion insulators~\cite{shafiei2022axion}, and high Chern number phases~\cite{shafiei2023high}, have garnered significant attention due to their diverse fundamental properties, applicable in next-generation spintronics~\cite{tokura2019magnetic}, and quantum computing~\cite{fan2016electric,che2020strongly}.
The giant spin-torque efficiency of magnetic TIs offers particular promise for developing energy-efficient spintronic devices. Extensive research efforts, both theoretical and experimental, have therefore been dedicated to exploring the fundamentals as well as potential applications of magnetic TIs~\cite{bernevig2022progress,otrokov2019prediction,he2019topological}. 

One common approach to realizing magnetic TIs involves the incorporation of magnetic dopants, such as in Bi$_2$Te$_3$ doped with Cr or V~\cite{tokura2019magnetic}. This method offers several advantages over alternative realizations of magnetic TIs, including the application of external magnetic fields~\cite{tokura2019magnetic}, proximity-induced magnetism from magnetic insulators~\cite{liu2023magnetic}, or the intrinsic magnetic TIs such as MnBi$_2$Te$_4$~\cite{he2020mnbi2te4}. It enables control over the direction and magnitude of magnetization induced by magnetic impurities~\cite{teng2019mn,zhang2018electronic}, while the fabrication techniques such as molecular beam epitaxy allow for tailored distribution of these impurities, locally or globally~\cite{tokura2019magnetic}. However, the introduction of magnetic dopants inevitably leads to disorder and impurity scattering during electronic transport. This magnetic disorder directly impacts the spin of electrons and, due to the strong intrinsic spin-orbit coupling, indirectly influences their momentum. Consequently, charge-current dissipation and backscattering are enhanced, limiting the appeal of these materials for practical applications.

To circumvent these limitations of magnetic TIs and foster their application in next-generation spintronic devices, it is crucial to identify strategies for realizing dissipationless states without backscattering. In this Letter, we reveal the possibility to control the electronic and transport properties of magnetic TIs by applying high-frequency linearly polarized light (LPL). More specifically, using the Floquet-Bloch theory, we demonstrate the states without backscattering in magnetic TIs through photo-induced resonances~\cite{cayssol2013floquet,mciver2012control}. The coupling between LPL and TI surface states is enabled through the presence of hexagonal warping in the Fermi surface. This coupling mechanism arises from the interplay between the Fermi surface deformation and the spin texture of the surface states, which is modulated by irradiation. The coupling of LPL with hexagonal warping alters the electronic band structure of the TI, which may effectively counteract the detrimental effects of magnetic impurities on the transport properties. Consequently, by controlling the amplitude and polarization angle $\theta$ of the incident light, it is possible to manipulate the conductance of the system. Note that in absence of hexagonal warping, the LPL-TI coupling vanishes. 

\paragraph{System and methods used} In this work, we consider the magnetic TI based on the Bi$_2$X$_3$ (X=Te,Se) material family doped with magnetic atoms. These materials are well known for their topological properties and rhombohedral crystal structure~\cite{zhang2009experimental}, with a unit cell comprising two Bi atoms and three chalcogen atoms in a quintuple layer~\cite{liu2010model,zhang2009topological}. When terminated on the (111) surface, these materials exhibit a distinctive electronic structure characterized by a uniform Dirac cone for low-energy states and a snowflake-like shape for high-energy states~\cite{liu2010model,chen2009experimental}. These bismuth-chalcogenide TIs are known to exhibit hexagonal warping of their Fermi surfaces~\cite{fu2009hexagonal,chen2009experimental}. This warping, stemming from the $C_{3v}$ rotational symmetry of these materials, introduces a $k$-cubic term into the band structure. This anisotropic term in momentum space enables anisotropic orbital coupling between the electron momentum and the electromagnetic radiation vector potential, and can induce interesting phenomena such as spin shift current~\cite{kim2017shift}. In presence of hexagonal warping the effective Hamiltonian describing the band structure of surface states near the $\Gamma$ point in the surface Brillouin zone becomes:
\begin{equation}\label{hamiltonian}
    H=\hbar\, v_F (k_x\sigma_y-k_y\sigma_x) + \lambda_h(k_x^3-3k_y^2\,k_x)\sigma_z + \Delta_h \sigma_z + M_{imp}\sigma_z,
\end{equation}
where $v_F = 2.9\times10^5$ ($4.0\times10^5$) m/s and $\lambda_h = 250$ ($128$) eV $\textup{\AA}^3$ stand for the Fermi velocity and hexagonal warping parameter for Bi$_2$Te$_3$~\cite{fu2009hexagonal} (Bi$_2$Se$_3$~\cite{kuroda2010hexagonally}), respectively, and $\sigma_{i=x,y,z}$ are the Pauli matrices. As the TI thickness is reduced, the surface states' wave functions on the top and bottom surfaces overlap, creating a gap at the Dirac point known as the hybridization gap~\cite{zhang2010crossover,shafiei2022controlling}, denoted by $\Delta_h$ in Eq.~\eqref{hamiltonian}. In our model for magnetically doped TI, we assume that magnetic impurities, such as Cr and V atoms, distributed over the lattice sites, contribute to a net out-of-plane magnetization $M_{imp}$, such that larger magnetization corresponds to a larger dopant concentration. The effect of magnetic impurities is incorporated as the Zeeman exchange in the last term of Eq.~\eqref{hamiltonian}~\cite{shafiei2023high}.

To analyze the impact of light on the magnetic TI, we employ Floquet theory. In presence of light with angular frequency $\omega$, one solves the time-dependent Schrödinger equation $i \hbar\, \partial_t \ket{\Psi_{\nu}(t)} = \operatorname{H}(t) \ket{\Psi_{\nu}(t)}$, where $\nu$ is the band index, and Hamiltonian is time-periodic, i.e. $H(t+T)=H(t)$, with $T=2\pi/\omega$. 
The Hamiltonian of the TI under irradiation can be described using the Peierls substitution ($\textbf{k} \to \textbf{k}+e\textbf{A}/\hbar$), where $\textbf{A}$ denotes the vector potential of the applied light~\cite{zhu2023floquet}.  
In this work, to prevent interband transitions, we strategically investigate the electromagnetic (high-frequency) spectral domain where the photon energy of the electromagnetic radiation is higher than the bandwidth of the TI.
We investigate TI surface irradiated perpendicularly with LPL, polarized at angle $\theta$ with respect to the $x$-axis (as shown in Fig.~\ref{fig:schematic_LPL}). This specific geometry is chosen to minimize contributions from linear and helicity-dependent photocurrents that can arise from oblique incidence~\cite{mciver2012control,shafiei2024towards}, enabling us to isolate the effects of interest. The vector potential of LPL is given by $\textbf{A}(t) = A_0 \cos (\omega t) [ \cos \theta \, \hat{x} + \sin \theta \, \hat{y} ]$. In the basis of Bloch waves, the time-dependent Hamiltonian is then expressed as \cite{xu2021light}:
\begin{equation}
    \Tilde{H}(\textbf{k},t) = \bra{m_k}\Tilde{H}(t)\ket{n_k} = H (\textbf{k}+\frac{e}{\hbar} \textbf{A}(t)).
\end{equation}
For the field $\textbf{A}(t)$ with period $T$, the Fourier transformation of $\Tilde{H}(\textbf{k},t)$ is:
\begin{equation}
    \Tilde{H}_m(\textbf{k}) = \frac{1}{T}  \int_{0}^{T} e^{i m \omega t}\, H(\textbf{k}+\frac{e}{\hbar} \textbf{A}(t))\,dt. 
\end{equation}
Using the van Vleck expansion, the effective Floquet Hamiltonian in the high-frequency regime can be obtained as \cite{mikami2016brillouin}:
\begin{eqnarray}\label{expansion}
\centering
H_F = H_0 + \sum_{m \neq 0} \frac{[H_{-m},H_m]}{2m\omega}\,+ \mathcal{O}(\omega^{-n}).
\end{eqnarray}
For convenience, in what follows the intensity of light $A_0$ will be given in reduced form as $\Tilde{A_0} = eA_0/\hbar$. The effective Floquet Hamiltonian for a magnetic TI using Eq.~\eqref{expansion} then becomes~\cite{choudhari2019effect}:
\begin{equation}\label{floq_hamiltonian}
    \begin{aligned}
        H_F& = \hbar v_F (k_x \sigma_y - k_y \sigma_x) + \lambda_h (k_x^3 - 3k_y^2 k_x) \sigma_z + \Delta_h \sigma_z \\
        &+ M_{imp} \sigma_z + \frac{3 \lambda_h \Tilde{A_0}^2}{2} (\cos{2\theta} k_x - \sin{2\theta} k_y) \sigma_z,
    \end{aligned}
\end{equation}
where the last term accounts for the coupling of LPL to the hexagonally warped Fermi surface.

\begin{figure}[t]
    \centering  
    \includegraphics[width=\linewidth]{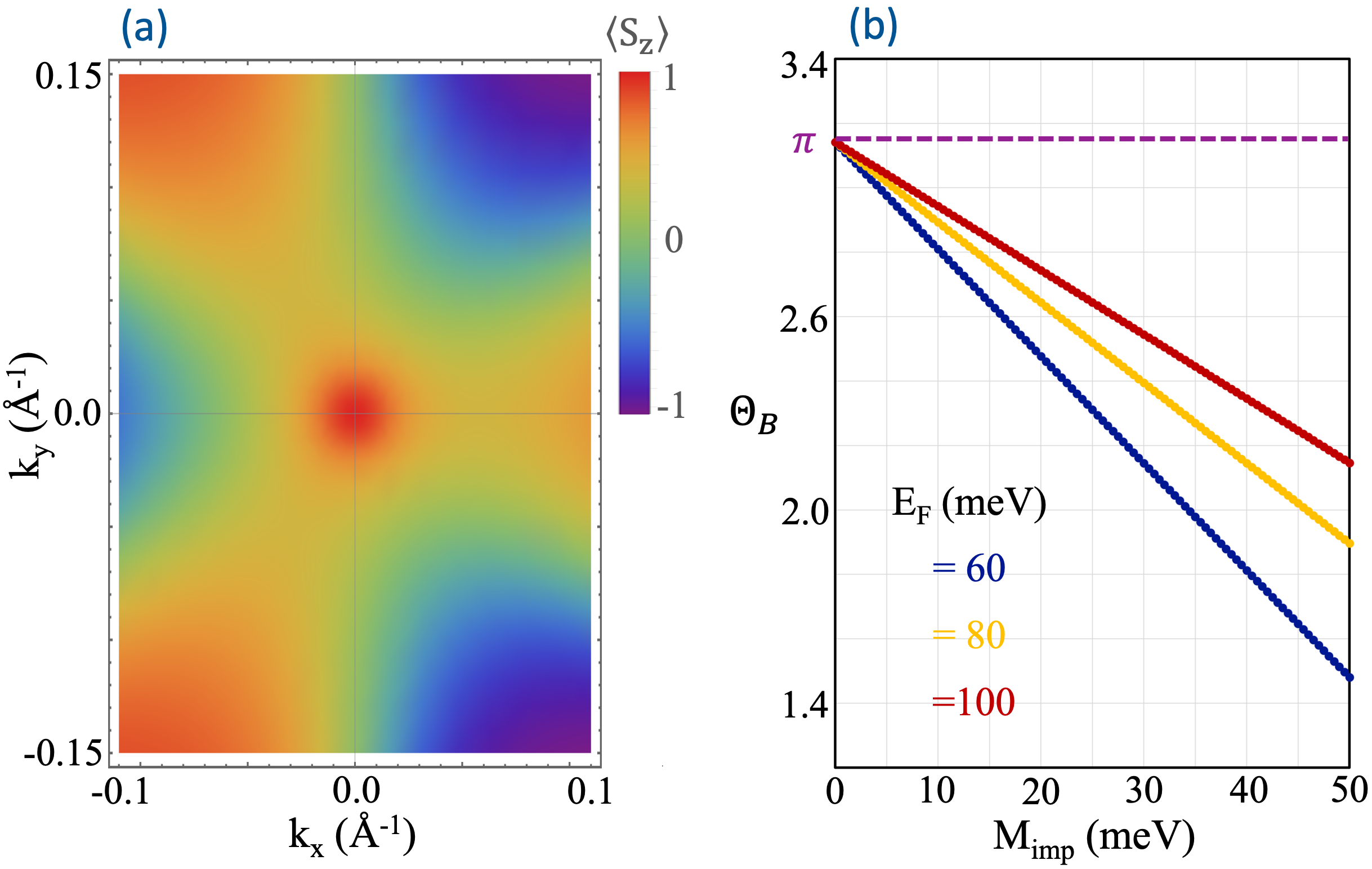}
    \caption{(a) $z$-component of the spin of surface-state electrons in magnetically doped Bi$_2$Te$_3$ with M$_{imp} = 30$~meV. The introduction of magnetic impurities breaks the perpendicular spin-momentum locking, resulting in a non-zero $z$-component of the spin. (b) Berry phase of surface states in magnetically doped Bi$_2$Te$_3$ as a function of magnetic impurity concentration (M$_{imp}$), for three different values of Fermi energy (E$_F$). Increasing M$_{imp}$ increases the deviation of the Berry phase from $\pi$, indicating enhanced backscattering.}
    \label{fig:BF_spin_magnetic}
\end{figure}

Finally, to investigate the transport properties of magnetically doped TIs, the Hamiltonian in momentum space [Eq.~\eqref{floq_hamiltonian}] can be represented in real space~\cite{nam2008electronic}. Employing a tight-binding approach, the Hamiltonian can be expressed in real space as:
\begin{equation}\label{real_hamiltonian}
    \begin{aligned}
        H_{R}&=\Sigma_{i} c_{i}^{\dagger} E_{on} c_{i} + \Sigma_{i,\bf{n_i}}(c_{i}^{\dagger} T_{\bf{n_i}} c_{i+\bf{n_i}}+H.c )\\
        &+ M_{imp}\sigma_z\otimes \sigma_0,
    \end{aligned}
\end{equation}
where $\bf{n_i} = \hat{x}, \hat{y}, \hat{z}$ are lattice vectors, and the operator $c_{i}^{\dagger}$ ($c_{i}$) creates (annihilates) an electron at site $i$. Further we have $E_{on}=(E_0 - 2\Sigma_{\bf{n_i}}B_{\bf{n_i}}) \sigma_{z} \otimes \sigma_{0}$, and $T_{\bf{n_i}}= B_{\bf{n_i}} \sigma_{z} \otimes \sigma_{0} - i (\frac{A_{\bf{n_i}}}{2}) \sigma_{x} \otimes \bf{\sigma} \cdot \bf{n_{i}}$, where $E_{on}$ and $T_{\bf{n_i}}$ stand for onsite energy and hopping parameters between unit cells, respectively, while T$_{\bf{n_i}}$ stands for the hopping parameters in the direction of these vectors. Fitting the band structure of this Hamiltonian with ab initio data for Bi$_2$Te$_3$~\cite{liu2010model,chu2011surface} yields the remaining parameters A$_\parallel$=0.5~eV, A$_z$=0.44~eV, E$_0$=0.3~eV, and B$_\parallel$=B$_z$=0.25~eV~\footnote{In Supplementary Material we show the corresponding calculations for Bi$_2$Se$_3$}. Finally, as in Eq.~\eqref{floq_hamiltonian}, to capture the coupling of LPL to the warped Fermi surface, terms $\frac{3 \lambda_h \Tilde{A_0}^2}{2} \cos{2\theta}\,\sigma_z$ and $\frac{3 \lambda_h \Tilde{A_0}^2}{2} \sin{2\theta}\, \sigma_z$ are added respectively to the hopping parameters along the $x$ and $y$ directions in the real-space Hamiltonian. The subsequent transport calculations were performed using the Landauer-B\"uttiker formalism. The conductance at zero temperature was calculated using Landauer's scattering method, where $G = \frac{e^2}{h}\,\sum_{n} T_n(E_F)$, with $T_n$ the probability of transmission in the $n$-th channel. In the case of a four-terminal setup, the conductance ($G$) can be defined in terms of the transmission coefficient between leads $i$ and $j$ as $T_{ij} = \trace [\Gamma_i G_{ij} \Gamma_j G_{ij}^{\dagger}]$, where G$_{ij}$ is the Green's function and $\Gamma_i$ is the coupling between the leads and the sample, defined in terms of self-energy $\Sigma_i$ as $\Gamma_i = i[\Sigma_i - \Sigma_i^{\dagger}]$~\cite{datta1997electronic}.

\begin{figure}[t]
    \centering     
    \includegraphics[width=0.968\linewidth]{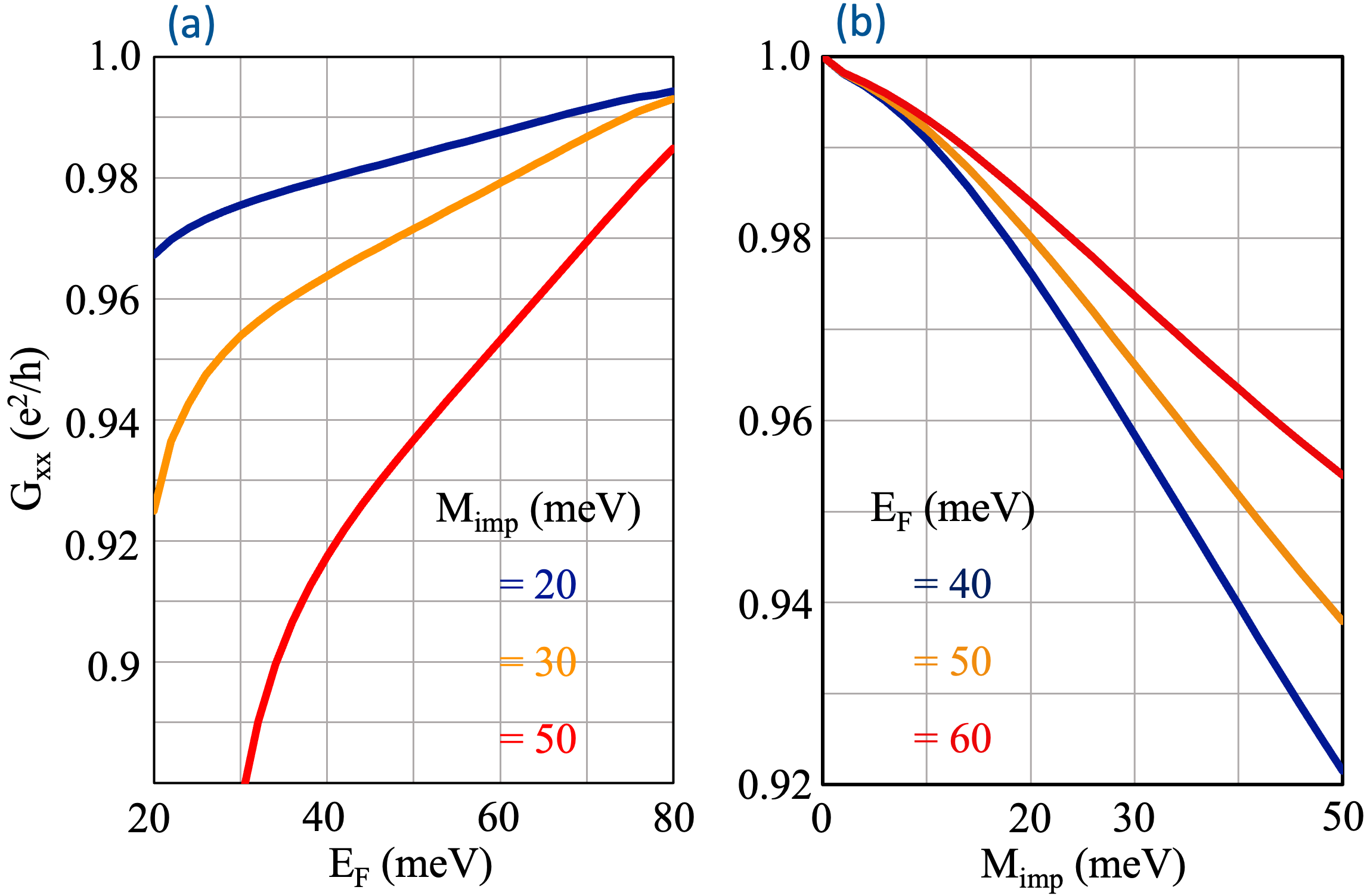}
    \caption{Longitudinal conductance (G$_{xx}$) of surface states in magnetically doped Bi$_2$Te$_3$ as a function of (a) Fermi energy and (b) net present magnetization. Increased magnetic impurity concentration leads to higher resistance, while increasing the Fermi energy reduces resistance.}
    \label{fig:transport_magnetic}
\end{figure}

\begin{figure*}[t]
    \centering     
    \includegraphics[width=0.98\linewidth]{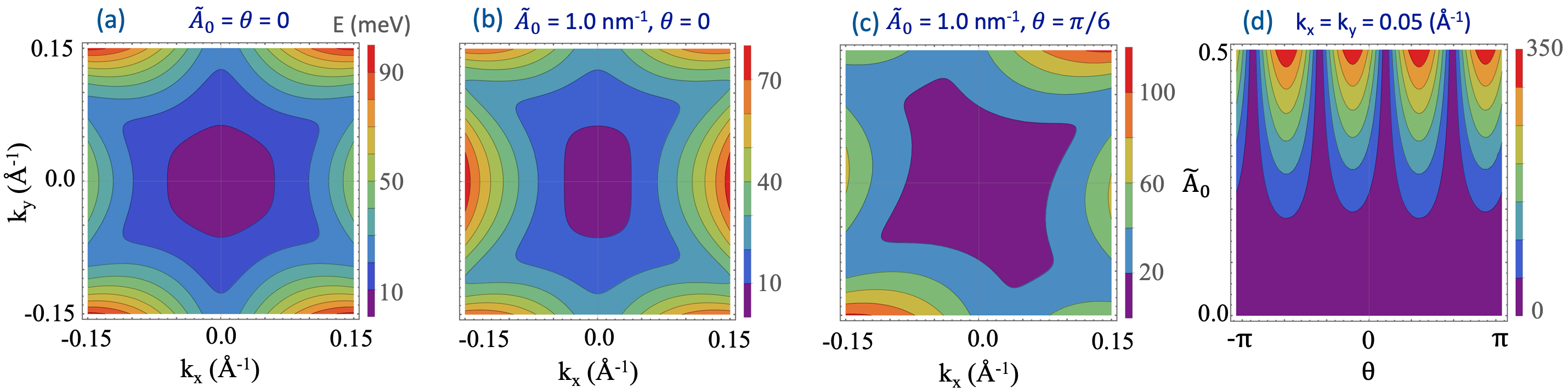}
    \caption{Contour plots of the surface band dispersion in momentum space of pristine Bi$_2$Te$_3$ for various intensities and polarization angles of applied LPL, considering the hexagonal warping of the Fermi surface. The effective Hamiltonian exhibits asymmetry under the transformation ($\theta \to -\theta$) and symmetry under ($\theta \to \theta + \pi$).}
    \label{fig:light_energy}
\end{figure*}

\paragraph{The backbone phenomena} As discussed previously, the introduction of magnetic dopants into TIs breaks TRS, opens a gap in the surface states, and disrupts spin-momentum locking. This results in a non-zero $z$-component of the spin of surface-state electrons, promoting backscattering. Figure~\ref{fig:BF_spin_magnetic}(a) shows the calculated change in the average $z$-component of the spin of surface-state electrons in the momentum space, in the case of magnetically doped Bi$_2$Te$_3$. In Fig.~\ref{fig:BF_spin_magnetic}(b) we show the calculated Berry phase, as a key indicator of backscattering strength. Indeed, the magnetic impurities destroy the helical surface states and the Berry phase deviates from its quantized value of $\pi$. A clear correlation is observed between the amount of magnetic impurities (total magnetization $M_{imp}$) and the deviation of the Berry phase from $\pi$. Increasing the impurity concentration thus results in enhanced backscattering and consequently higher resistivity in the system.

Next we address the transport properties of surface states in magnetic TIs. Figure~\ref{fig:transport_magnetic} shows the calculated longitudinal conductance (G$_{xx}$) of surface states in magnetically doped Bi$_2$Te$_3$ for various Fermi energies (Fig.~\ref{fig:transport_magnetic}(a)) and magnetization strengths (Fig.~\ref{fig:transport_magnetic}(b)). These results show that increasing the concentration of magnetic impurities leads to a significant deviation of the longitudinal conductance from the quantized value of $e^2/h$, indicating increased dissipation within the system. Conversely, increasing the Fermi energy for a fixed magnetic impurity concentration can reduce the resistance of surface states, consistent with the Berry phase results shown in Fig.~\ref{fig:BF_spin_magnetic}.

Let us now switch to the properties of the system under illumination. To date, it was observed that applying high-frequency circularly polarized light (CPL) to a TI results in the opening of a gap on the surface states, even in the absence of hexagonal warping, by generating a time reversal breaking mass term~\cite{xu2021light,shafiei2024floquet}. In contrast to CPL, high-frequency LPL does not induce a gap in the surface states of TIs unless there is hexagonal warping in the Fermi surface of the TI~\cite{choudhari2019effect}. We therefore first validate the effect of LPL on the surface state dispersion of Bi$_2$Te$_3$, after properly accounting for its hexagonal warping. Notably, the band structure and surface state properties can indeed be controlled by adjusting both the intensity and the polarization angle of the applied light, as shown in Figure~\ref{fig:light_energy}. 
The application of LPL in this case results in an effective Hamiltonian that is asymmetric under the transformation $\theta \to -\theta$ and symmetric under the transformation $\theta \to \theta + \pi$, as seen in the latter figure.

\paragraph{Control of resistivity by light} To this point, we have explained the main effects of magnetic doping of TIs, which leads to backscattering, the destruction of helical surface states, and increased dissipation. We have also explained the impact of LPL on TI surface states, considering hexagonal warping of the Fermi surface of the material. Next, we combine all these ingredients to pursue dissipationless charge transport in magnetically doped TIs using high-frequency LPL. 

\begin{figure*}[t]
   \centering
   \includegraphics[width=0.99\linewidth]{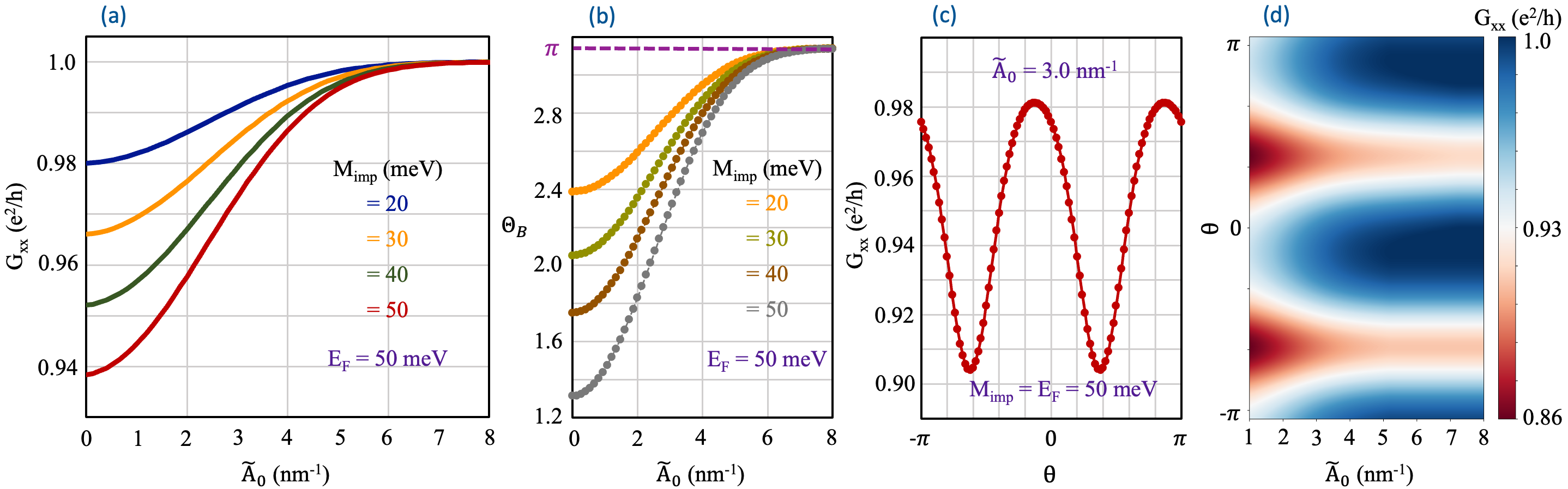}
   \caption{(a) Longitudinal conductance ($G_{xx}$) of the magnetically doped Bi$_2$Te$_3$ with orthogonally incident LPL, for $E_F = 50$ meV and different magnetization strengths, as a function of the applied light intensity (while $\theta=0$). Increasing the LPL intensity reduces the resistance, ultimately leading to dissipationless charge transport. (b) Berry phase of a magnetic TI under linearly polarized light. Increasing the light intensity reduces backscattering and surface state resistance. (c) Conductance as a function of the light polarization angle $\theta$, for other parameters as given in the figure. (d) Contour plot of conductance change as a function of intensity and polarization angle of the incident light. }
    \label{fig:transport_LPL}
\end{figure*}

Considering magnetically doped Bi$_2$Te$_3$ and accounting for hexagonal warping, we present in Fig.~\ref{fig:transport_LPL}(a) our calculations for the longitudinal transport of surface states. The data are shown for different magnetic impurity concentrations and a range of irradiation intensities, for fixed polarization $\theta$ = 0 and a Fermi energy of 50 meV. These results clearly demonstrate that increasing LPL intensity reduces dissipation, ultimately leading to an \textit{entirely dissipationless charge transport channel} in the system. It is important to note that the coupling strength between LPL and TI is directly proportional to the magnitude of the hexagonal warping of the surface states. A less pronounced hexagonal warping leads to weaker light-matter coupling, such that dissipationless states can be realized only at higher light intensities. Our calculations indicate that, comparing Bi$_2$Te$_3$ to Bi$_2$Se$_3$, the approximately twice larger hexagonal warping in Bi$_2$Te$_3$ would necessitate a lower light intensity of $\Tilde{A_0} \approx  7.0$ nm$^{-1}$, as opposed to $\Tilde{A_0} \approx  10.0$ nm$^{-1}$ for Bi$_2$Se$_3$, to achieve dissipationless states.

As an additional benchmark, we also calculated the Berry phase of a magnetically doped TI under LPL illumination. These calculations, corresponding to the longitudinal conductance calculations of the surface state shown in Fig.~\ref{fig:transport_LPL}(a), were performed for various magnetization strengths at a fixed Fermi energy of E$_F$=50~meV. The results, plotted in Fig.~\ref{fig:transport_LPL}(b), show that increasing the light intensity reduces the deviation of the Berry phase from its quantized value of $\pi$, indicating a decrease in backscattering and, consequently, decrease of dissipation in the system.

Last but not least, we reveal that the transport properties of the system can be controlled not only by adjusting the light intensity but also by modifying the polarization angle ($\theta$). 
We present in Fig.~\ref{fig:transport_LPL}(c) the calculated longitudinal conductance of the surface state as a function of the polarization angle of the applied LPL, for fixed values of light intensity ($\Tilde{A_0}$ = 3.0 nm$^{-1}$), magnetization (M$_{imp}$ = 50 meV), and Fermi energy (E$_F$ = 50 meV). This figure clearly shows that longitudinal transport exhibits a distinct symmetry under transformation $\theta \to \theta+\pi$. This observed symmetry aligns with the anticipated symmetry of the Hamiltonian for this specific transformation. The shift of the maximum conductance away from $\theta$ = 0 is directly attributed to the angular dependence of the hopping parameters within the system's Hamiltonian, stemming from the LPL coupling to the warping of the Fermi surface. Fig.~\ref{fig:transport_LPL}(d) summarizes the calculated conductance as a function of the intensity and polarization angle of the light (for M$_{imp}$ = 50 meV and E$_F$ = 50 meV), showing the large range of conductance variation (10-15\%) promising for optical sensing purposes.

\paragraph{Conclusion} In summary, we have revealed the potential of linearly polarized light (LPL) to suppress backscattering induced by magnetic impurities in magnetically doped topological insulators (TIs) with hexagonal warping. Unlike circularly polarized light, which can open a gap in the Dirac cone, LPL can couple to surface states in the presence of hexagonal warping, providing a mechanism to control the electronic and transport properties of TIs. By employing Floquet theory and comprehensive Berry phase and electronic transport calculations using real-space Hamiltonian parametrized from first principles, we have shown that such control can be realized in a wide range yet precisely, as a function of the amplitude and polarization angle of the incident light. Conversely, the measured resistance can serve for advanced optical sensing with respect to the incident irradiation. We have applied our analysis to realistic cases of magnetically doped Bi$_2$Te$_3$ and Bi$_2$Se$_3$ (see Supplementary Material), as readily available and studied experimentally, and confirmed that strategic application of LPL can indeed significantly reduce or even completely eliminate backscattering, enhancing the carrier mobility and diminishing dissipation in these materials, paving the way for a wide range of energy-efficient electronic, spintronic, and sensing devices.

The authors thank Farhad Fazileh for useful discussions. This research was supported by the Research Foundation-Flanders (FWO-Vlaanderen) and the Special Research Funds (BOF) of the University of Antwerp.

\bibliographystyle{apsrev4-1}
\bibliography{manuscript}

\begin{thebibliography}{40}%
\makeatletter
\providecommand \@ifxundefined [1]{%
 \@ifx{#1\undefined}
}%
\providecommand \@ifnum [1]{%
 \ifnum #1\expandafter \@firstoftwo
 \else \expandafter \@secondoftwo
 \fi
}%
\providecommand \@ifx [1]{%
 \ifx #1\expandafter \@firstoftwo
 \else \expandafter \@secondoftwo
 \fi
}%
\providecommand \natexlab [1]{#1}%
\providecommand \enquote  [1]{``#1''}%
\providecommand \bibnamefont  [1]{#1}%
\providecommand \bibfnamefont [1]{#1}%
\providecommand \citenamefont [1]{#1}%
\providecommand \href@noop [0]{\@secondoftwo}%
\providecommand \href [0]{\begingroup \@sanitize@url \@href}%
\providecommand \@href[1]{\@@startlink{#1}\@@href}%
\providecommand \@@href[1]{\endgroup#1\@@endlink}%
\providecommand \@sanitize@url [0]{\catcode `\\12\catcode `\$12\catcode `\&12\catcode `\#12\catcode `\^12\catcode `\_12\catcode `\%12\relax}%
\providecommand \@@startlink[1]{}%
\providecommand \@@endlink[0]{}%
\providecommand \url  [0]{\begingroup\@sanitize@url \@url }%
\providecommand \@url [1]{\endgroup\@href {#1}{\urlprefix }}%
\providecommand \urlprefix  [0]{URL }%
\providecommand \Eprint [0]{\href }%
\providecommand \doibase [0]{http://dx.doi.org/}%
\providecommand \selectlanguage [0]{\@gobble}%
\providecommand \bibinfo  [0]{\@secondoftwo}%
\providecommand \bibfield  [0]{\@secondoftwo}%
\providecommand \translation [1]{[#1]}%
\providecommand \BibitemOpen [0]{}%
\providecommand \bibitemStop [0]{}%
\providecommand \bibitemNoStop [0]{.\EOS\space}%
\providecommand \EOS [0]{\spacefactor3000\relax}%
\providecommand \BibitemShut  [1]{\csname bibitem#1\endcsname}%
\let\auto@bib@innerbib\@empty
\bibitem [{\citenamefont {Fu}\ \emph {et~al.}(2007)\citenamefont {Fu}, \citenamefont {Kane},\ and\ \citenamefont {Mele}}]{fu2007topological}%
  \BibitemOpen
  \bibfield  {author} {\bibinfo {author} {\bibfnamefont {L.}~\bibnamefont {Fu}}, \bibinfo {author} {\bibfnamefont {C.~L.}\ \bibnamefont {Kane}}, \ and\ \bibinfo {author} {\bibfnamefont {E.~J.}\ \bibnamefont {Mele}},\ }\href@noop {} {\bibfield  {journal} {\bibinfo  {journal} {Physical Review Letters}\ }\textbf {\bibinfo {volume} {98}},\ \bibinfo {pages} {106803} (\bibinfo {year} {2007})}\BibitemShut {NoStop}%
\bibitem [{\citenamefont {Hasan}\ and\ \citenamefont {Kane}(2010)}]{hasan2010colloquium}%
  \BibitemOpen
  \bibfield  {author} {\bibinfo {author} {\bibfnamefont {M.~Z.}\ \bibnamefont {Hasan}}\ and\ \bibinfo {author} {\bibfnamefont {C.~L.}\ \bibnamefont {Kane}},\ }\href@noop {} {\bibfield  {journal} {\bibinfo  {journal} {Reviews of Modern Physics}\ }\textbf {\bibinfo {volume} {82}},\ \bibinfo {pages} {3045} (\bibinfo {year} {2010})}\BibitemShut {NoStop}%
\bibitem [{\citenamefont {Qi}\ and\ \citenamefont {Zhang}(2011)}]{qi2011topological}%
  \BibitemOpen
  \bibfield  {author} {\bibinfo {author} {\bibfnamefont {X.-L.}\ \bibnamefont {Qi}}\ and\ \bibinfo {author} {\bibfnamefont {S.-C.}\ \bibnamefont {Zhang}},\ }\href@noop {} {\bibfield  {journal} {\bibinfo  {journal} {Reviews of Modern Physics}\ }\textbf {\bibinfo {volume} {83}},\ \bibinfo {pages} {1057} (\bibinfo {year} {2011})}\BibitemShut {NoStop}%
\bibitem [{\citenamefont {Hasan}\ and\ \citenamefont {Moore}(2011)}]{hasan2011three}%
  \BibitemOpen
  \bibfield  {author} {\bibinfo {author} {\bibfnamefont {M.~Z.}\ \bibnamefont {Hasan}}\ and\ \bibinfo {author} {\bibfnamefont {J.~E.}\ \bibnamefont {Moore}},\ }\href@noop {} {\bibfield  {journal} {\bibinfo  {journal} {Annu. Rev. Condens. Matter Phys.}\ }\textbf {\bibinfo {volume} {2}},\ \bibinfo {pages} {55} (\bibinfo {year} {2011})}\BibitemShut {NoStop}%
\bibitem [{\citenamefont {Tian}\ \emph {et~al.}(2017)\citenamefont {Tian}, \citenamefont {Yu}, \citenamefont {Shi},\ and\ \citenamefont {Wang}}]{tian2017property}%
  \BibitemOpen
  \bibfield  {author} {\bibinfo {author} {\bibfnamefont {W.}~\bibnamefont {Tian}}, \bibinfo {author} {\bibfnamefont {W.}~\bibnamefont {Yu}}, \bibinfo {author} {\bibfnamefont {J.}~\bibnamefont {Shi}}, \ and\ \bibinfo {author} {\bibfnamefont {Y.}~\bibnamefont {Wang}},\ }\href@noop {} {\bibfield  {journal} {\bibinfo  {journal} {Materials}\ }\textbf {\bibinfo {volume} {10}},\ \bibinfo {pages} {814} (\bibinfo {year} {2017})}\BibitemShut {NoStop}%
\bibitem [{\citenamefont {SHEN}(2018)}]{shun2018topological}%
  \BibitemOpen
  \bibfield  {author} {\bibinfo {author} {\bibfnamefont {S.-Q.}\ \bibnamefont {SHEN}},\ }\href@noop {} {\emph {\bibinfo {title} {Topological Insulators: Dirac Equation in Condensed Matter}}}\ (\bibinfo  {publisher} {Springer},\ \bibinfo {year} {2018})\BibitemShut {NoStop}%
\bibitem [{\citenamefont {Tokura}\ \emph {et~al.}(2019)\citenamefont {Tokura}, \citenamefont {Yasuda},\ and\ \citenamefont {Tsukazaki}}]{tokura2019magnetic}%
  \BibitemOpen
  \bibfield  {author} {\bibinfo {author} {\bibfnamefont {Y.}~\bibnamefont {Tokura}}, \bibinfo {author} {\bibfnamefont {K.}~\bibnamefont {Yasuda}}, \ and\ \bibinfo {author} {\bibfnamefont {A.}~\bibnamefont {Tsukazaki}},\ }\href@noop {} {\bibfield  {journal} {\bibinfo  {journal} {Nature Reviews Physics}\ }\textbf {\bibinfo {volume} {1}},\ \bibinfo {pages} {126} (\bibinfo {year} {2019})}\BibitemShut {NoStop}%
\bibitem [{\citenamefont {Chang}\ \emph {et~al.}(2013)\citenamefont {Chang}, \citenamefont {Zhang}, \citenamefont {Feng}, \citenamefont {Shen}, \citenamefont {Zhang}, \citenamefont {Guo}, \citenamefont {Li}, \citenamefont {Ou}, \citenamefont {Wei}, \citenamefont {Wang} \emph {et~al.}}]{chang2013experimental}%
  \BibitemOpen
  \bibfield  {author} {\bibinfo {author} {\bibfnamefont {C.-Z.}\ \bibnamefont {Chang}}, \bibinfo {author} {\bibfnamefont {J.}~\bibnamefont {Zhang}}, \bibinfo {author} {\bibfnamefont {X.}~\bibnamefont {Feng}}, \bibinfo {author} {\bibfnamefont {J.}~\bibnamefont {Shen}}, \bibinfo {author} {\bibfnamefont {Z.}~\bibnamefont {Zhang}}, \bibinfo {author} {\bibfnamefont {M.}~\bibnamefont {Guo}}, \bibinfo {author} {\bibfnamefont {K.}~\bibnamefont {Li}}, \bibinfo {author} {\bibfnamefont {Y.}~\bibnamefont {Ou}}, \bibinfo {author} {\bibfnamefont {P.}~\bibnamefont {Wei}}, \bibinfo {author} {\bibfnamefont {L.-L.}\ \bibnamefont {Wang}},  \emph {et~al.},\ }\href@noop {} {\bibfield  {journal} {\bibinfo  {journal} {Science}\ }\textbf {\bibinfo {volume} {340}},\ \bibinfo {pages} {167} (\bibinfo {year} {2013})}\BibitemShut {NoStop}%
\bibitem [{\citenamefont {Shafiei}\ \emph {et~al.}(2022{\natexlab{a}})\citenamefont {Shafiei}, \citenamefont {Fazileh}, \citenamefont {Peeters},\ and\ \citenamefont {Milo{\v{s}}evi{\'c}}}]{shafiei2022axion}%
  \BibitemOpen
  \bibfield  {author} {\bibinfo {author} {\bibfnamefont {M.}~\bibnamefont {Shafiei}}, \bibinfo {author} {\bibfnamefont {F.}~\bibnamefont {Fazileh}}, \bibinfo {author} {\bibfnamefont {F.~M.}\ \bibnamefont {Peeters}}, \ and\ \bibinfo {author} {\bibfnamefont {M.~V.}\ \bibnamefont {Milo{\v{s}}evi{\'c}}},\ }\href@noop {} {\bibfield  {journal} {\bibinfo  {journal} {Physical Review Materials}\ }\textbf {\bibinfo {volume} {6}},\ \bibinfo {pages} {074205} (\bibinfo {year} {2022}{\natexlab{a}})}\BibitemShut {NoStop}%
\bibitem [{\citenamefont {Shafiei}\ \emph {et~al.}(2023)\citenamefont {Shafiei}, \citenamefont {Fazileh}, \citenamefont {Peeters},\ and\ \citenamefont {Milo{\v{s}}evi{\'c}}}]{shafiei2023high}%
  \BibitemOpen
  \bibfield  {author} {\bibinfo {author} {\bibfnamefont {M.}~\bibnamefont {Shafiei}}, \bibinfo {author} {\bibfnamefont {F.}~\bibnamefont {Fazileh}}, \bibinfo {author} {\bibfnamefont {F.~M.}\ \bibnamefont {Peeters}}, \ and\ \bibinfo {author} {\bibfnamefont {M.~V.}\ \bibnamefont {Milo{\v{s}}evi{\'c}}},\ }\href@noop {} {\bibfield  {journal} {\bibinfo  {journal} {Physical Review B}\ }\textbf {\bibinfo {volume} {107}},\ \bibinfo {pages} {195119} (\bibinfo {year} {2023})}\BibitemShut {NoStop}%
\bibitem [{\citenamefont {Fan}\ \emph {et~al.}(2016)\citenamefont {Fan}, \citenamefont {Kou}, \citenamefont {Upadhyaya}, \citenamefont {Shao}, \citenamefont {Pan}, \citenamefont {Lang}, \citenamefont {Che}, \citenamefont {Tang}, \citenamefont {Montazeri}, \citenamefont {Murata} \emph {et~al.}}]{fan2016electric}%
  \BibitemOpen
  \bibfield  {author} {\bibinfo {author} {\bibfnamefont {Y.}~\bibnamefont {Fan}}, \bibinfo {author} {\bibfnamefont {X.}~\bibnamefont {Kou}}, \bibinfo {author} {\bibfnamefont {P.}~\bibnamefont {Upadhyaya}}, \bibinfo {author} {\bibfnamefont {Q.}~\bibnamefont {Shao}}, \bibinfo {author} {\bibfnamefont {L.}~\bibnamefont {Pan}}, \bibinfo {author} {\bibfnamefont {M.}~\bibnamefont {Lang}}, \bibinfo {author} {\bibfnamefont {X.}~\bibnamefont {Che}}, \bibinfo {author} {\bibfnamefont {J.}~\bibnamefont {Tang}}, \bibinfo {author} {\bibfnamefont {M.}~\bibnamefont {Montazeri}}, \bibinfo {author} {\bibfnamefont {K.}~\bibnamefont {Murata}},  \emph {et~al.},\ }\href@noop {} {\bibfield  {journal} {\bibinfo  {journal} {Nature Nanotechnology}\ }\textbf {\bibinfo {volume} {11}},\ \bibinfo {pages} {352} (\bibinfo {year} {2016})}\BibitemShut {NoStop}%
\bibitem [{\citenamefont {Che}\ \emph {et~al.}(2020)\citenamefont {Che}, \citenamefont {Pan}, \citenamefont {Vareskic}, \citenamefont {Zou}, \citenamefont {Pan}, \citenamefont {Zhang}, \citenamefont {Yin}, \citenamefont {Wu}, \citenamefont {Shao}, \citenamefont {Deng} \emph {et~al.}}]{che2020strongly}%
  \BibitemOpen
  \bibfield  {author} {\bibinfo {author} {\bibfnamefont {X.}~\bibnamefont {Che}}, \bibinfo {author} {\bibfnamefont {Q.}~\bibnamefont {Pan}}, \bibinfo {author} {\bibfnamefont {B.}~\bibnamefont {Vareskic}}, \bibinfo {author} {\bibfnamefont {J.}~\bibnamefont {Zou}}, \bibinfo {author} {\bibfnamefont {L.}~\bibnamefont {Pan}}, \bibinfo {author} {\bibfnamefont {P.}~\bibnamefont {Zhang}}, \bibinfo {author} {\bibfnamefont {G.}~\bibnamefont {Yin}}, \bibinfo {author} {\bibfnamefont {H.}~\bibnamefont {Wu}}, \bibinfo {author} {\bibfnamefont {Q.}~\bibnamefont {Shao}}, \bibinfo {author} {\bibfnamefont {P.}~\bibnamefont {Deng}},  \emph {et~al.},\ }\href@noop {} {\bibfield  {journal} {\bibinfo  {journal} {Advanced Materials}\ }\textbf {\bibinfo {volume} {32}},\ \bibinfo {pages} {1907661} (\bibinfo {year} {2020})}\BibitemShut {NoStop}%
\bibitem [{\citenamefont {Bernevig}\ \emph {et~al.}(2022)\citenamefont {Bernevig}, \citenamefont {Felser},\ and\ \citenamefont {Beidenkopf}}]{bernevig2022progress}%
  \BibitemOpen
  \bibfield  {author} {\bibinfo {author} {\bibfnamefont {B.~A.}\ \bibnamefont {Bernevig}}, \bibinfo {author} {\bibfnamefont {C.}~\bibnamefont {Felser}}, \ and\ \bibinfo {author} {\bibfnamefont {H.}~\bibnamefont {Beidenkopf}},\ }\href@noop {} {\bibfield  {journal} {\bibinfo  {journal} {Nature}\ }\textbf {\bibinfo {volume} {603}},\ \bibinfo {pages} {41} (\bibinfo {year} {2022})}\BibitemShut {NoStop}%
\bibitem [{\citenamefont {Otrokov}\ \emph {et~al.}(2019)\citenamefont {Otrokov}, \citenamefont {Klimovskikh}, \citenamefont {Bentmann}, \citenamefont {Estyunin}, \citenamefont {Zeugner}, \citenamefont {Aliev}, \citenamefont {Ga{\ss}}, \citenamefont {Wolter}, \citenamefont {Koroleva}, \citenamefont {Shikin} \emph {et~al.}}]{otrokov2019prediction}%
  \BibitemOpen
  \bibfield  {author} {\bibinfo {author} {\bibfnamefont {M.~M.}\ \bibnamefont {Otrokov}}, \bibinfo {author} {\bibfnamefont {I.~I.}\ \bibnamefont {Klimovskikh}}, \bibinfo {author} {\bibfnamefont {H.}~\bibnamefont {Bentmann}}, \bibinfo {author} {\bibfnamefont {D.}~\bibnamefont {Estyunin}}, \bibinfo {author} {\bibfnamefont {A.}~\bibnamefont {Zeugner}}, \bibinfo {author} {\bibfnamefont {Z.~S.}\ \bibnamefont {Aliev}}, \bibinfo {author} {\bibfnamefont {S.}~\bibnamefont {Ga{\ss}}}, \bibinfo {author} {\bibfnamefont {A.}~\bibnamefont {Wolter}}, \bibinfo {author} {\bibfnamefont {A.}~\bibnamefont {Koroleva}}, \bibinfo {author} {\bibfnamefont {A.~M.}\ \bibnamefont {Shikin}},  \emph {et~al.},\ }\href@noop {} {\bibfield  {journal} {\bibinfo  {journal} {Nature}\ }\textbf {\bibinfo {volume} {576}},\ \bibinfo {pages} {416} (\bibinfo {year} {2019})}\BibitemShut {NoStop}%
\bibitem [{\citenamefont {He}\ \emph {et~al.}(2019)\citenamefont {He}, \citenamefont {Sun},\ and\ \citenamefont {He}}]{he2019topological}%
  \BibitemOpen
  \bibfield  {author} {\bibinfo {author} {\bibfnamefont {M.}~\bibnamefont {He}}, \bibinfo {author} {\bibfnamefont {H.}~\bibnamefont {Sun}}, \ and\ \bibinfo {author} {\bibfnamefont {Q.~L.}\ \bibnamefont {He}},\ }\href@noop {} {\bibfield  {journal} {\bibinfo  {journal} {Frontiers of Physics}\ }\textbf {\bibinfo {volume} {14}},\ \bibinfo {pages} {1} (\bibinfo {year} {2019})}\BibitemShut {NoStop}%
\bibitem [{\citenamefont {Liu}\ and\ \citenamefont {Hesjedal}(2023)}]{liu2023magnetic}%
  \BibitemOpen
  \bibfield  {author} {\bibinfo {author} {\bibfnamefont {J.}~\bibnamefont {Liu}}\ and\ \bibinfo {author} {\bibfnamefont {T.}~\bibnamefont {Hesjedal}},\ }\href@noop {} {\bibfield  {journal} {\bibinfo  {journal} {Advanced Materials}\ }\textbf {\bibinfo {volume} {35}},\ \bibinfo {pages} {2102427} (\bibinfo {year} {2023})}\BibitemShut {NoStop}%
\bibitem [{\citenamefont {He}(2020)}]{he2020mnbi2te4}%
  \BibitemOpen
  \bibfield  {author} {\bibinfo {author} {\bibfnamefont {K.}~\bibnamefont {He}},\ }\href@noop {} {\bibfield  {journal} {\bibinfo  {journal} {npj Quantum Materials}\ }\textbf {\bibinfo {volume} {5}},\ \bibinfo {pages} {90} (\bibinfo {year} {2020})}\BibitemShut {NoStop}%
\bibitem [{\citenamefont {Teng}\ \emph {et~al.}(2019)\citenamefont {Teng}, \citenamefont {Liu},\ and\ \citenamefont {Li}}]{teng2019mn}%
  \BibitemOpen
  \bibfield  {author} {\bibinfo {author} {\bibfnamefont {J.}~\bibnamefont {Teng}}, \bibinfo {author} {\bibfnamefont {N.}~\bibnamefont {Liu}}, \ and\ \bibinfo {author} {\bibfnamefont {Y.}~\bibnamefont {Li}},\ }\href@noop {} {\bibfield  {journal} {\bibinfo  {journal} {Journal of Semiconductors}\ }\textbf {\bibinfo {volume} {40}},\ \bibinfo {pages} {081507} (\bibinfo {year} {2019})}\BibitemShut {NoStop}%
\bibitem [{\citenamefont {Zhang}\ \emph {et~al.}(2018)\citenamefont {Zhang}, \citenamefont {West}, \citenamefont {Lee}, \citenamefont {Qiu}, \citenamefont {Chang}, \citenamefont {Moodera}, \citenamefont {San~Hor}, \citenamefont {Zhang},\ and\ \citenamefont {Wu}}]{zhang2018electronic}%
  \BibitemOpen
  \bibfield  {author} {\bibinfo {author} {\bibfnamefont {W.}~\bibnamefont {Zhang}}, \bibinfo {author} {\bibfnamefont {D.}~\bibnamefont {West}}, \bibinfo {author} {\bibfnamefont {S.~H.}\ \bibnamefont {Lee}}, \bibinfo {author} {\bibfnamefont {Y.}~\bibnamefont {Qiu}}, \bibinfo {author} {\bibfnamefont {C.-Z.}\ \bibnamefont {Chang}}, \bibinfo {author} {\bibfnamefont {J.~S.}\ \bibnamefont {Moodera}}, \bibinfo {author} {\bibfnamefont {Y.}~\bibnamefont {San~Hor}}, \bibinfo {author} {\bibfnamefont {S.}~\bibnamefont {Zhang}}, \ and\ \bibinfo {author} {\bibfnamefont {W.}~\bibnamefont {Wu}},\ }\href@noop {} {\bibfield  {journal} {\bibinfo  {journal} {Physical Review B}\ }\textbf {\bibinfo {volume} {98}},\ \bibinfo {pages} {115165} (\bibinfo {year} {2018})}\BibitemShut {NoStop}%
\bibitem [{\citenamefont {Cayssol}\ \emph {et~al.}(2013)\citenamefont {Cayssol}, \citenamefont {D{\'o}ra}, \citenamefont {Simon},\ and\ \citenamefont {Moessner}}]{cayssol2013floquet}%
  \BibitemOpen
  \bibfield  {author} {\bibinfo {author} {\bibfnamefont {J.}~\bibnamefont {Cayssol}}, \bibinfo {author} {\bibfnamefont {B.}~\bibnamefont {D{\'o}ra}}, \bibinfo {author} {\bibfnamefont {F.}~\bibnamefont {Simon}}, \ and\ \bibinfo {author} {\bibfnamefont {R.}~\bibnamefont {Moessner}},\ }\href@noop {} {\bibfield  {journal} {\bibinfo  {journal} {physica status solidi (RRL)--Rapid Research Letters}\ }\textbf {\bibinfo {volume} {7}},\ \bibinfo {pages} {101} (\bibinfo {year} {2013})}\BibitemShut {NoStop}%
\bibitem [{\citenamefont {McIver}\ \emph {et~al.}(2012)\citenamefont {McIver}, \citenamefont {Hsieh}, \citenamefont {Steinberg}, \citenamefont {Jarillo-Herrero},\ and\ \citenamefont {Gedik}}]{mciver2012control}%
  \BibitemOpen
  \bibfield  {author} {\bibinfo {author} {\bibfnamefont {J.}~\bibnamefont {McIver}}, \bibinfo {author} {\bibfnamefont {D.}~\bibnamefont {Hsieh}}, \bibinfo {author} {\bibfnamefont {H.}~\bibnamefont {Steinberg}}, \bibinfo {author} {\bibfnamefont {P.}~\bibnamefont {Jarillo-Herrero}}, \ and\ \bibinfo {author} {\bibfnamefont {N.}~\bibnamefont {Gedik}},\ }\href@noop {} {\bibfield  {journal} {\bibinfo  {journal} {Nature nanotechnology}\ }\textbf {\bibinfo {volume} {7}},\ \bibinfo {pages} {96} (\bibinfo {year} {2012})}\BibitemShut {NoStop}%
\bibitem [{\citenamefont {Zhang}\ \emph {et~al.}(2009{\natexlab{a}})\citenamefont {Zhang}, \citenamefont {Cheng}, \citenamefont {Chen}, \citenamefont {Jia}, \citenamefont {Ma}, \citenamefont {He}, \citenamefont {Wang}, \citenamefont {Zhang}, \citenamefont {Dai}, \citenamefont {Fang} \emph {et~al.}}]{zhang2009experimental}%
  \BibitemOpen
  \bibfield  {author} {\bibinfo {author} {\bibfnamefont {T.}~\bibnamefont {Zhang}}, \bibinfo {author} {\bibfnamefont {P.}~\bibnamefont {Cheng}}, \bibinfo {author} {\bibfnamefont {X.}~\bibnamefont {Chen}}, \bibinfo {author} {\bibfnamefont {J.-F.}\ \bibnamefont {Jia}}, \bibinfo {author} {\bibfnamefont {X.}~\bibnamefont {Ma}}, \bibinfo {author} {\bibfnamefont {K.}~\bibnamefont {He}}, \bibinfo {author} {\bibfnamefont {L.}~\bibnamefont {Wang}}, \bibinfo {author} {\bibfnamefont {H.}~\bibnamefont {Zhang}}, \bibinfo {author} {\bibfnamefont {X.}~\bibnamefont {Dai}}, \bibinfo {author} {\bibfnamefont {Z.}~\bibnamefont {Fang}},  \emph {et~al.},\ }\href@noop {} {\bibfield  {journal} {\bibinfo  {journal} {Physical Review Letters}\ }\textbf {\bibinfo {volume} {103}},\ \bibinfo {pages} {266803} (\bibinfo {year} {2009}{\natexlab{a}})}\BibitemShut {NoStop}%
\bibitem [{\citenamefont {Liu}\ \emph {et~al.}(2010)\citenamefont {Liu}, \citenamefont {Qi}, \citenamefont {Zhang}, \citenamefont {Dai}, \citenamefont {Fang},\ and\ \citenamefont {Zhang}}]{liu2010model}%
  \BibitemOpen
  \bibfield  {author} {\bibinfo {author} {\bibfnamefont {C.-X.}\ \bibnamefont {Liu}}, \bibinfo {author} {\bibfnamefont {X.-L.}\ \bibnamefont {Qi}}, \bibinfo {author} {\bibfnamefont {H.}~\bibnamefont {Zhang}}, \bibinfo {author} {\bibfnamefont {X.}~\bibnamefont {Dai}}, \bibinfo {author} {\bibfnamefont {Z.}~\bibnamefont {Fang}}, \ and\ \bibinfo {author} {\bibfnamefont {S.-C.}\ \bibnamefont {Zhang}},\ }\href@noop {} {\bibfield  {journal} {\bibinfo  {journal} {Physical Review B}\ }\textbf {\bibinfo {volume} {82}},\ \bibinfo {pages} {045122} (\bibinfo {year} {2010})}\BibitemShut {NoStop}%
\bibitem [{\citenamefont {Zhang}\ \emph {et~al.}(2009{\natexlab{b}})\citenamefont {Zhang}, \citenamefont {Liu}, \citenamefont {Qi}, \citenamefont {Dai}, \citenamefont {Fang},\ and\ \citenamefont {Zhang}}]{zhang2009topological}%
  \BibitemOpen
  \bibfield  {author} {\bibinfo {author} {\bibfnamefont {H.}~\bibnamefont {Zhang}}, \bibinfo {author} {\bibfnamefont {C.-X.}\ \bibnamefont {Liu}}, \bibinfo {author} {\bibfnamefont {X.-L.}\ \bibnamefont {Qi}}, \bibinfo {author} {\bibfnamefont {X.}~\bibnamefont {Dai}}, \bibinfo {author} {\bibfnamefont {Z.}~\bibnamefont {Fang}}, \ and\ \bibinfo {author} {\bibfnamefont {S.-C.}\ \bibnamefont {Zhang}},\ }\href@noop {} {\bibfield  {journal} {\bibinfo  {journal} {Nature Physics}\ }\textbf {\bibinfo {volume} {5}},\ \bibinfo {pages} {438} (\bibinfo {year} {2009}{\natexlab{b}})}\BibitemShut {NoStop}%
\bibitem [{\citenamefont {Chen}\ \emph {et~al.}(2009)\citenamefont {Chen}, \citenamefont {Analytis}, \citenamefont {Chu}, \citenamefont {Liu}, \citenamefont {Mo}, \citenamefont {Qi}, \citenamefont {Zhang}, \citenamefont {Lu}, \citenamefont {Dai}, \citenamefont {Fang} \emph {et~al.}}]{chen2009experimental}%
  \BibitemOpen
  \bibfield  {author} {\bibinfo {author} {\bibfnamefont {Y.}~\bibnamefont {Chen}}, \bibinfo {author} {\bibfnamefont {J.~G.}\ \bibnamefont {Analytis}}, \bibinfo {author} {\bibfnamefont {J.-H.}\ \bibnamefont {Chu}}, \bibinfo {author} {\bibfnamefont {Z.}~\bibnamefont {Liu}}, \bibinfo {author} {\bibfnamefont {S.-K.}\ \bibnamefont {Mo}}, \bibinfo {author} {\bibfnamefont {X.-L.}\ \bibnamefont {Qi}}, \bibinfo {author} {\bibfnamefont {H.}~\bibnamefont {Zhang}}, \bibinfo {author} {\bibfnamefont {D.}~\bibnamefont {Lu}}, \bibinfo {author} {\bibfnamefont {X.}~\bibnamefont {Dai}}, \bibinfo {author} {\bibfnamefont {Z.}~\bibnamefont {Fang}},  \emph {et~al.},\ }\href@noop {} {\bibfield  {journal} {\bibinfo  {journal} {Science}\ }\textbf {\bibinfo {volume} {325}},\ \bibinfo {pages} {178} (\bibinfo {year} {2009})}\BibitemShut {NoStop}%
\bibitem [{\citenamefont {Fu}(2009)}]{fu2009hexagonal}%
  \BibitemOpen
  \bibfield  {author} {\bibinfo {author} {\bibfnamefont {L.}~\bibnamefont {Fu}},\ }\href@noop {} {\bibfield  {journal} {\bibinfo  {journal} {Physical Review Letters}\ }\textbf {\bibinfo {volume} {103}},\ \bibinfo {pages} {266801} (\bibinfo {year} {2009})}\BibitemShut {NoStop}%
\bibitem [{\citenamefont {Kim}\ \emph {et~al.}(2017)\citenamefont {Kim}, \citenamefont {Morimoto},\ and\ \citenamefont {Nagaosa}}]{kim2017shift}%
  \BibitemOpen
  \bibfield  {author} {\bibinfo {author} {\bibfnamefont {K.~W.}\ \bibnamefont {Kim}}, \bibinfo {author} {\bibfnamefont {T.}~\bibnamefont {Morimoto}}, \ and\ \bibinfo {author} {\bibfnamefont {N.}~\bibnamefont {Nagaosa}},\ }\href@noop {} {\bibfield  {journal} {\bibinfo  {journal} {Physical Review B}\ }\textbf {\bibinfo {volume} {95}},\ \bibinfo {pages} {035134} (\bibinfo {year} {2017})}\BibitemShut {NoStop}%
\bibitem [{\citenamefont {Kuroda}\ \emph {et~al.}(2010)\citenamefont {Kuroda}, \citenamefont {Arita}, \citenamefont {Miyamoto}, \citenamefont {Ye}, \citenamefont {Jiang}, \citenamefont {Kimura}, \citenamefont {Krasovskii}, \citenamefont {Chulkov}, \citenamefont {Iwasawa}, \citenamefont {Okuda} \emph {et~al.}}]{kuroda2010hexagonally}%
  \BibitemOpen
  \bibfield  {author} {\bibinfo {author} {\bibfnamefont {K.}~\bibnamefont {Kuroda}}, \bibinfo {author} {\bibfnamefont {M.}~\bibnamefont {Arita}}, \bibinfo {author} {\bibfnamefont {K.}~\bibnamefont {Miyamoto}}, \bibinfo {author} {\bibfnamefont {M.}~\bibnamefont {Ye}}, \bibinfo {author} {\bibfnamefont {J.}~\bibnamefont {Jiang}}, \bibinfo {author} {\bibfnamefont {A.}~\bibnamefont {Kimura}}, \bibinfo {author} {\bibfnamefont {E.}~\bibnamefont {Krasovskii}}, \bibinfo {author} {\bibfnamefont {E.}~\bibnamefont {Chulkov}}, \bibinfo {author} {\bibfnamefont {H.}~\bibnamefont {Iwasawa}}, \bibinfo {author} {\bibfnamefont {T.}~\bibnamefont {Okuda}},  \emph {et~al.},\ }\href@noop {} {\bibfield  {journal} {\bibinfo  {journal} {Physical Review Letters}\ }\textbf {\bibinfo {volume} {105}},\ \bibinfo {pages} {076802} (\bibinfo {year} {2010})}\BibitemShut {NoStop}%
\bibitem [{\citenamefont {Zhang}\ \emph {et~al.}(2010)\citenamefont {Zhang}, \citenamefont {He}, \citenamefont {Chang}, \citenamefont {Song}, \citenamefont {Wang}, \citenamefont {Chen}, \citenamefont {Jia}, \citenamefont {Fang}, \citenamefont {Dai}, \citenamefont {Shan} \emph {et~al.}}]{zhang2010crossover}%
  \BibitemOpen
  \bibfield  {author} {\bibinfo {author} {\bibfnamefont {Y.}~\bibnamefont {Zhang}}, \bibinfo {author} {\bibfnamefont {K.}~\bibnamefont {He}}, \bibinfo {author} {\bibfnamefont {C.-Z.}\ \bibnamefont {Chang}}, \bibinfo {author} {\bibfnamefont {C.-L.}\ \bibnamefont {Song}}, \bibinfo {author} {\bibfnamefont {L.-L.}\ \bibnamefont {Wang}}, \bibinfo {author} {\bibfnamefont {X.}~\bibnamefont {Chen}}, \bibinfo {author} {\bibfnamefont {J.-F.}\ \bibnamefont {Jia}}, \bibinfo {author} {\bibfnamefont {Z.}~\bibnamefont {Fang}}, \bibinfo {author} {\bibfnamefont {X.}~\bibnamefont {Dai}}, \bibinfo {author} {\bibfnamefont {W.-Y.}\ \bibnamefont {Shan}},  \emph {et~al.},\ }\href@noop {} {\bibfield  {journal} {\bibinfo  {journal} {Nature Physics}\ }\textbf {\bibinfo {volume} {6}},\ \bibinfo {pages} {584} (\bibinfo {year} {2010})}\BibitemShut {NoStop}%
\bibitem [{\citenamefont {Shafiei}\ \emph {et~al.}(2022{\natexlab{b}})\citenamefont {Shafiei}, \citenamefont {Fazileh}, \citenamefont {Peeters},\ and\ \citenamefont {Milo{\v{s}}evi{\'c}}}]{shafiei2022controlling}%
  \BibitemOpen
  \bibfield  {author} {\bibinfo {author} {\bibfnamefont {M.}~\bibnamefont {Shafiei}}, \bibinfo {author} {\bibfnamefont {F.}~\bibnamefont {Fazileh}}, \bibinfo {author} {\bibfnamefont {F.~M.}\ \bibnamefont {Peeters}}, \ and\ \bibinfo {author} {\bibfnamefont {M.~V.}\ \bibnamefont {Milo{\v{s}}evi{\'c}}},\ }\href@noop {} {\bibfield  {journal} {\bibinfo  {journal} {Physical Review B}\ }\textbf {\bibinfo {volume} {106}},\ \bibinfo {pages} {035119} (\bibinfo {year} {2022}{\natexlab{b}})}\BibitemShut {NoStop}%
\bibitem [{\citenamefont {Zhu}\ \emph {et~al.}(2023)\citenamefont {Zhu}, \citenamefont {Wang},\ and\ \citenamefont {Zhang}}]{zhu2023floquet}%
  \BibitemOpen
  \bibfield  {author} {\bibinfo {author} {\bibfnamefont {T.}~\bibnamefont {Zhu}}, \bibinfo {author} {\bibfnamefont {H.}~\bibnamefont {Wang}}, \ and\ \bibinfo {author} {\bibfnamefont {H.}~\bibnamefont {Zhang}},\ }\href@noop {} {\bibfield  {journal} {\bibinfo  {journal} {Physical Review B}\ }\textbf {\bibinfo {volume} {107}},\ \bibinfo {pages} {085151} (\bibinfo {year} {2023})}\BibitemShut {NoStop}%
\bibitem [{\citenamefont {Shafiei}\ \emph {et~al.}(2024{\natexlab{a}})\citenamefont {Shafiei}, \citenamefont {Moayeri},\ and\ \citenamefont {Milo{\v{s}}evi{\'c}}}]{shafiei2024towards}%
  \BibitemOpen
  \bibfield  {author} {\bibinfo {author} {\bibfnamefont {M.}~\bibnamefont {Shafiei}}, \bibinfo {author} {\bibfnamefont {S.~S.}\ \bibnamefont {Moayeri}}, \ and\ \bibinfo {author} {\bibfnamefont {M.~V.}\ \bibnamefont {Milo{\v{s}}evi{\'c}}},\ }\href@noop {} {\bibfield  {journal} {\bibinfo  {journal} {arXiv preprint arXiv:2411.12040}\ } (\bibinfo {year} {2024}{\natexlab{a}})}\BibitemShut {NoStop}%
\bibitem [{\citenamefont {Xu}\ \emph {et~al.}(2021)\citenamefont {Xu}, \citenamefont {Zhou},\ and\ \citenamefont {Li}}]{xu2021light}%
  \BibitemOpen
  \bibfield  {author} {\bibinfo {author} {\bibfnamefont {H.}~\bibnamefont {Xu}}, \bibinfo {author} {\bibfnamefont {J.}~\bibnamefont {Zhou}}, \ and\ \bibinfo {author} {\bibfnamefont {J.}~\bibnamefont {Li}},\ }\href@noop {} {\bibfield  {journal} {\bibinfo  {journal} {Advanced Science}\ }\textbf {\bibinfo {volume} {8}},\ \bibinfo {pages} {2101508} (\bibinfo {year} {2021})}\BibitemShut {NoStop}%
\bibitem [{\citenamefont {Mikami}\ \emph {et~al.}(2016)\citenamefont {Mikami}, \citenamefont {Kitamura}, \citenamefont {Yasuda}, \citenamefont {Tsuji}, \citenamefont {Oka},\ and\ \citenamefont {Aoki}}]{mikami2016brillouin}%
  \BibitemOpen
  \bibfield  {author} {\bibinfo {author} {\bibfnamefont {T.}~\bibnamefont {Mikami}}, \bibinfo {author} {\bibfnamefont {S.}~\bibnamefont {Kitamura}}, \bibinfo {author} {\bibfnamefont {K.}~\bibnamefont {Yasuda}}, \bibinfo {author} {\bibfnamefont {N.}~\bibnamefont {Tsuji}}, \bibinfo {author} {\bibfnamefont {T.}~\bibnamefont {Oka}}, \ and\ \bibinfo {author} {\bibfnamefont {H.}~\bibnamefont {Aoki}},\ }\href@noop {} {\bibfield  {journal} {\bibinfo  {journal} {Physical Review B}\ }\textbf {\bibinfo {volume} {93}},\ \bibinfo {pages} {144307} (\bibinfo {year} {2016})}\BibitemShut {NoStop}%
\bibitem [{\citenamefont {Choudhari}\ and\ \citenamefont {Deo}(2019)}]{choudhari2019effect}%
  \BibitemOpen
  \bibfield  {author} {\bibinfo {author} {\bibfnamefont {T.}~\bibnamefont {Choudhari}}\ and\ \bibinfo {author} {\bibfnamefont {N.}~\bibnamefont {Deo}},\ }\href@noop {} {\bibfield  {journal} {\bibinfo  {journal} {Physical Review B}\ }\textbf {\bibinfo {volume} {100}},\ \bibinfo {pages} {035303} (\bibinfo {year} {2019})}\BibitemShut {NoStop}%
\bibitem [{\citenamefont {Nam~Do}\ \emph {et~al.}(2008)\citenamefont {Nam~Do}, \citenamefont {Nguyen}, \citenamefont {Dollfus},\ and\ \citenamefont {Bournel}}]{nam2008electronic}%
  \BibitemOpen
  \bibfield  {author} {\bibinfo {author} {\bibfnamefont {V.}~\bibnamefont {Nam~Do}}, \bibinfo {author} {\bibfnamefont {V.~H.}\ \bibnamefont {Nguyen}}, \bibinfo {author} {\bibfnamefont {P.}~\bibnamefont {Dollfus}}, \ and\ \bibinfo {author} {\bibfnamefont {A.}~\bibnamefont {Bournel}},\ }\href@noop {} {\bibfield  {journal} {\bibinfo  {journal} {Journal of Applied Physics}\ }\textbf {\bibinfo {volume} {104}} (\bibinfo {year} {2008})}\BibitemShut {NoStop}%
\bibitem [{\citenamefont {Chu}\ \emph {et~al.}(2011)\citenamefont {Chu}, \citenamefont {Shi},\ and\ \citenamefont {Shen}}]{chu2011surface}%
  \BibitemOpen
  \bibfield  {author} {\bibinfo {author} {\bibfnamefont {R.-L.}\ \bibnamefont {Chu}}, \bibinfo {author} {\bibfnamefont {J.}~\bibnamefont {Shi}}, \ and\ \bibinfo {author} {\bibfnamefont {S.-Q.}\ \bibnamefont {Shen}},\ }\href@noop {} {\bibfield  {journal} {\bibinfo  {journal} {Physical Review B—Condensed Matter and Materials Physics}\ }\textbf {\bibinfo {volume} {84}},\ \bibinfo {pages} {085312} (\bibinfo {year} {2011})}\BibitemShut {NoStop}%
\bibitem [{Note1()}]{Note1}%
  \BibitemOpen
  \bibinfo {note} {In Supplementary Material we show the corresponding calculations for Bi$_2$Se$_3$}\BibitemShut {NoStop}%
\bibitem [{\citenamefont {Datta}(1997)}]{datta1997electronic}%
  \BibitemOpen
  \bibfield  {author} {\bibinfo {author} {\bibfnamefont {S.}~\bibnamefont {Datta}},\ }\href@noop {} {\emph {\bibinfo {title} {Electronic transport in mesoscopic systems}}}\ (\bibinfo  {publisher} {Cambridge University Press},\ \bibinfo {year} {1997})\BibitemShut {NoStop}%
\bibitem [{\citenamefont {Shafiei}\ \emph {et~al.}(2024{\natexlab{b}})\citenamefont {Shafiei}, \citenamefont {Fazileh}, \citenamefont {Peeters},\ and\ \citenamefont {Milo{\v{s}}evi{\'c}}}]{shafiei2024floquet}%
  \BibitemOpen
  \bibfield  {author} {\bibinfo {author} {\bibfnamefont {M.}~\bibnamefont {Shafiei}}, \bibinfo {author} {\bibfnamefont {F.}~\bibnamefont {Fazileh}}, \bibinfo {author} {\bibfnamefont {F.~M.}\ \bibnamefont {Peeters}}, \ and\ \bibinfo {author} {\bibfnamefont {M.~V.}\ \bibnamefont {Milo{\v{s}}evi{\'c}}},\ }\href@noop {} {\bibfield  {journal} {\bibinfo  {journal} {SciPost Physics Core}\ }\textbf {\bibinfo {volume} {7}},\ \bibinfo {pages} {024} (\bibinfo {year} {2024}{\natexlab{b}})}\BibitemShut {NoStop}%
\end{thebibliography}%
					
\end {document}